\documentclass{aastex631}

\shorttitle{GRB~221009A}
\shortauthors{Frederiks et al.}
\def \GRB {GRB~221009A}
\def \KW  {Konus-\textit{WIND}}

\begin{document}

\title{Properties of the Extremely Energetic GRB~221009A from Konus-\textit{WIND} and \textit{SRG}/ART-XC Observations}

\correspondingauthor{D.~Frederiks}
\email{fred@mail.ioffe.ru, ddfrederiks@gmail.com}

\author{D.~Frederiks}
\author{D.~Svinkin}
\author{A.~L.~Lysenko}
\affiliation{Ioffe Institute, 26 Politekhnicheskaya, St Petersburg, 194021, Russia}
\author{S. Molkov}
\affiliation{Space Research Institute, 	Russian Academy of Sciences, Profsoyuznaya 84/32, 117997 Moscow, Russia}
\author{A.~Tsvetkova}
\author{M.~Ulanov}
\author{A.~Ridnaia}
\affiliation{Ioffe Institute, 26 Politekhnicheskaya, St Petersburg, 194021, Russia}
\author{A.~A.~Lutovinov}
\author{I.~Lapshov}
\author{A.~Tkachenko}
\author{V.~Levin}
\affiliation{Space Research Institute, 	Russian Academy of Sciences, Profsoyuznaya 84/32, 117997 Moscow, Russia}

\begin{abstract}
We report on \KW\ (KW) and {\it Mikhail Pavlinsky}~ART-XC telescope observations 
and analysis of a nearby \GRB\,, the brightest $\gamma$-ray burst (GRB) 
detected by KW for $>$28 years of observations. 
The pulsed prompt phase of the burst emission lasts for $\sim 600$~s 
and is followed by a steady power-law decay lasting for more than 25~ks. 
From the analysis of the KW and ART-XC light curves and the KW spectral data 
we derive time-averaged spectral peak energy of the burst $E_p\approx 2.6$~MeV,
$E_p$ at the brightest emission peak $\approx 3.0$~MeV, 
the total 20~keV--10~MeV energy fluence of $\approx0.22$~erg~cm$^{-2}$, 
and the peak energy flux in the same band of $\approx 0.031$~erg~cm$^{-2}$~s$^{-1}$.
The enormous observed fluence and peak flux imply, at redshift $z=0.151$, 
huge values of isotropic energy release $E_{\mathrm{iso}}\approx1.2\times10^{55}$~erg 
(or $\gtrsim 6.5$~solar rest mass) and isotropic peak luminosity 
$L_{\mathrm{iso}}\approx3.4\times10^{54}$~erg~s$^{-1}$ (64~ms scale), 
making \GRB\ the most energetic and one of the most luminous bursts 
observed since the beginning of the GRB cosmological era in 1997. 
The isotropic energetics of the burst fit nicely both ``Amati'' and ``Yonetoku'' 
hardness-intensity correlations for $>$300~KW long GRBs, 
implying that \GRB\ is most likely a very hard, 
super-energetic version of a ``normal'' long GRB.

\end{abstract}


\section{Introduction} \label{sec:intro}

Cosmological gamma-ray bursts (GRBs) are thought to be produced by at least two distinct classes of catastrophic events: 
mergers of binary compact objects, such as two neutron stars or a neutron star and a black hole, typically produce short, 
$\lesssim 2$~s, so-called Type~I GRBs; the core collapse of massive stars produce typically long (Type~II) GRBs. 
See, e.g., \citet{Zhang_2009ApJ_703_1696} for more information on the Type~I/II classification scheme.

GRBs have been the target of many observational efforts at all wavelengths, from a multitude of space- and ground-based observatories (see \citealt{Tsvetkova_2022Univ....8..373T} for a recent GRB facility review). GRBs occur at a rate of about 1 day$^{-1}$ and, with many thousand events observed to date  
\citep{Mazets_1981Ap&SS..80....3M,Briggs_1996ApJ...459...40B,Atteia_1999A&AS..138..421A,Frontera_2009ApJS..180..192F,Guidorzi_2011A&A...526A..49G,Goldstein_2013ApJS..208...21G,Lien_2016ApJ...829....7L,Svinkin_2016ApJS..224...10S,Kozlova_2019JPhCS1400b2014K}, the basic properties of their prompt $\gamma$-ray emissions are well established.
The bursts last from a fraction of a second to several thousand seconds, showing a wide range of structures in their light curves and having a typical peak energy 
in the 100~keV--1~MeV range. The overall observed GRB fluences range from $10^{-7}$ to as high as $10^{-3}$~erg~cm$^{-2}$.

The GRB cosmological origin was established about 25 years ago and it became clear that the observed flux corresponds to an enormous emitted energy, making GRBs the most luminous objects in the sky.
The measured GRB isotropic-equivalent energy release $E_\mathrm{iso}$ and isotropic peak luminosity $L_\mathrm{iso}$ have broad distributions
\citep{Amati_2002A&A...390...81A,Amati_2008MNRAS.391..577A,Yonetoku_2004ApJ...609..935Y,Gruber_2011A&A...531A..20G,Atteia_2017ApJ...837..119A,Tsvetkova_2017ApJ...850..161T,Tsvetkova_2021ApJ...908...83T}
and tend to follow a number of empirical correlations between rest-frame parameters of GRB prompt emission,
e.g., the ``Amati'' \citep{Amati_2002A&A...390...81A}, ``Yonetoku'' \citep{Yonetoku_2004ApJ...609..935Y} and ``Ghirlanda'' \citep{Ghirlanda_2007A&A...466..127G} relations. 
The most intense GRBs reaching close to $E_\mathrm{iso}\sim 10^{55}$~erg \citep{Abdo_2009Sci...323.1688A,Greiner_2009A&A...498...89G,Tsvetkova_2017ApJ...850..161T} 
and $L_\mathrm{iso} \sim 5\times 10^{54}$~erg~s$^{-1}$ \citep{Frederiks_2013ApJ...779..151F,Svinkin_2021GCN.30276....1S}.
An upper limit on GRB isotropic energy has recently been predicted ($E_\mathrm{iso}\sim 3.8\times10^{54}$~erg, \citealt{Dado_2022ApJ...940L...4D}), 
which, together with a strong cutoff of the $E_\mathrm{iso}$ distribution above $1-3\times10^{54}$~erg,
suggested from the analysis of \KW\ and \textit{Fermi}-GBM samples of GRBs with known 
redshifts \citep{Atteia_2017ApJ...837..119A,Tsvetkova_2017ApJ...850..161T,Tsvetkova_2021ApJ...908...83T},  
imply very rare detections of extremely energetic GRBs. Bright, nearby GRBs provide a unique opportunity to probe the central-engine physics, 
prompt emission and afterglow emission mechanisms, as well as the GRB local environment. 
So far, only a few such bursts have been observed.

On 2022 October 9 at about 13:17:00 UTC, an extremely intense \GRB\ 
was detected by many space-based missions:   
\textit{Fermi} (GBM and LAT; \citealt{2022GCN.32636....1V, 2022GCN.32642....1L,2022GCN.32637....1B,2022GCN.32658....1P}),
Konus-\textit{Wind}~\citep{2022GCN.32641....1S, 2022GCN.32668....1F}, 
\textit{AGILE}~(MCAL and GRID; \citealt{2022GCN.32650....1U, 2022ATel15662....1P}), 
\textit{INTEGRAL} (SPI-ACS; \citealt{2022GCN.32660....1G}), 
\textit{Insight}-HXMT~\citep{2022ATel15660....1T}, 
\textit{Solar Orbiter} (STIX; \citealt{2022GCN.32661....1X}),
\textit{Spektr-RG} (ART-XC; \citealt{2022GCN.32663....1L}), 
\textit{GRBAlpha}~\citep{2022GCN.32685....1R}, 
\textit{SIRI-2}~\citep{2022GCN.32746....1M},
\textit{GECAM-C}~\citep{2022GCN.32751....1L}, 
and \textit{BepiColombo} (MGNS; \citealt{2022GCN.32805....1K}).
The initial analysis of the burst showed that the prompt emission was so intense that it saturated almost all instruments.

About 53 minutes later, the bright hard X-ray and optical afterglow, 
initially designated as a transient Swift~J$1913.1+1946$, was detected by 
the Neil Gehrels Swift Observatory (Burst Alert Telescope, BAT; X-Ray Telescope, XRT; 
and Ultraviolet/Optical Telescope, UVOT; \citealt{2022GCN.32632....1D, 2022GCN.32688....1K}).
The multiwavelength follow-up observations led to detection of bright optical afterglow and 
spectroscopic redshift determination of $z=0.151$, which implies a luminosity distance $d_L$ of 745~Mpc
\footnote{Assuming a flat $\Lambda$CDM cosmology with $H_0 = 67.8$~km~s$^{-1}$~Mpc$^{-1}$ $\Omega_M = 0.315$, 
\citep{Planck_2020A&A...641A...6P}}
 \citep{2022GCN.32648....1D,2022GCN.32686....1C,Malesani_2023arXiv230207891M}
The possible supernova associated with the burst was discovered a few days after 
the GRB~\citep{2022GCN.32769....1B, 2022GCN.32800....1D, 2022GCN.32818....1B}.

A preliminary analysis of the \KW\,(KW) detection revealed that \GRB\ 
is the most intense $\gamma$-ray burst observed by the instrument  \citep{2022GCN.32668....1F}.
The brightness of the main burst episode did not allow to perform the standard KW spectral 
analysis of the emission. However, with preliminary dead-time corrections applied, 
a rough estimate of the energy fluence of the 600-s long burst was obtained ($\sim0.052$~erg~cm$^{-2}$), 
which is the highest value observed for GRBs for 28 years of the KW operation.

The {\it Mikhail Pavlinsky} ART-XC telescope (ART-XC) observed \GRB\ outside its field of view (FoV).
The burst signal passed through the telescope's side shield, but it was clearly visible in all seven detectors. 
A preliminary analysis showed that the GRB light curve has a complex shape which can be restored with 
good accuracy \citep{2022GCN.32663....1L}. 

In this work, we present the detailed analysis of KW and ART-XC detections of \GRB.
Both instruments operate in interplanetary space, in orbits around Lagrange points L1 and L2,
respectively, that allowed to observe the burst for its whole duration 
in stable background conditions.
From the KW temporal and spectral data corrected for instrumental effects and 
the ART-XC light curve data we derive key parameters of \GRB\ prompt emission 
in the observer frame, estimate the event energetics in the cosmological rest frame of the source, 
and discuss this extraordinary burst in the context of the KW sample of long GRBs. 

Throughout the paper all errors reported are 90\% conf. levels unless otherwise specified.
We adopt the conventional notation $Q_k=Q/10^k$ and use cgs units unless otherwise noted.

\section{Observations} \label{sec:obs}
\subsection{Konus-Wind} \label{sec:obsKW}

\begin{figure*}
	\centering
	\includegraphics[width=0.9\textwidth]{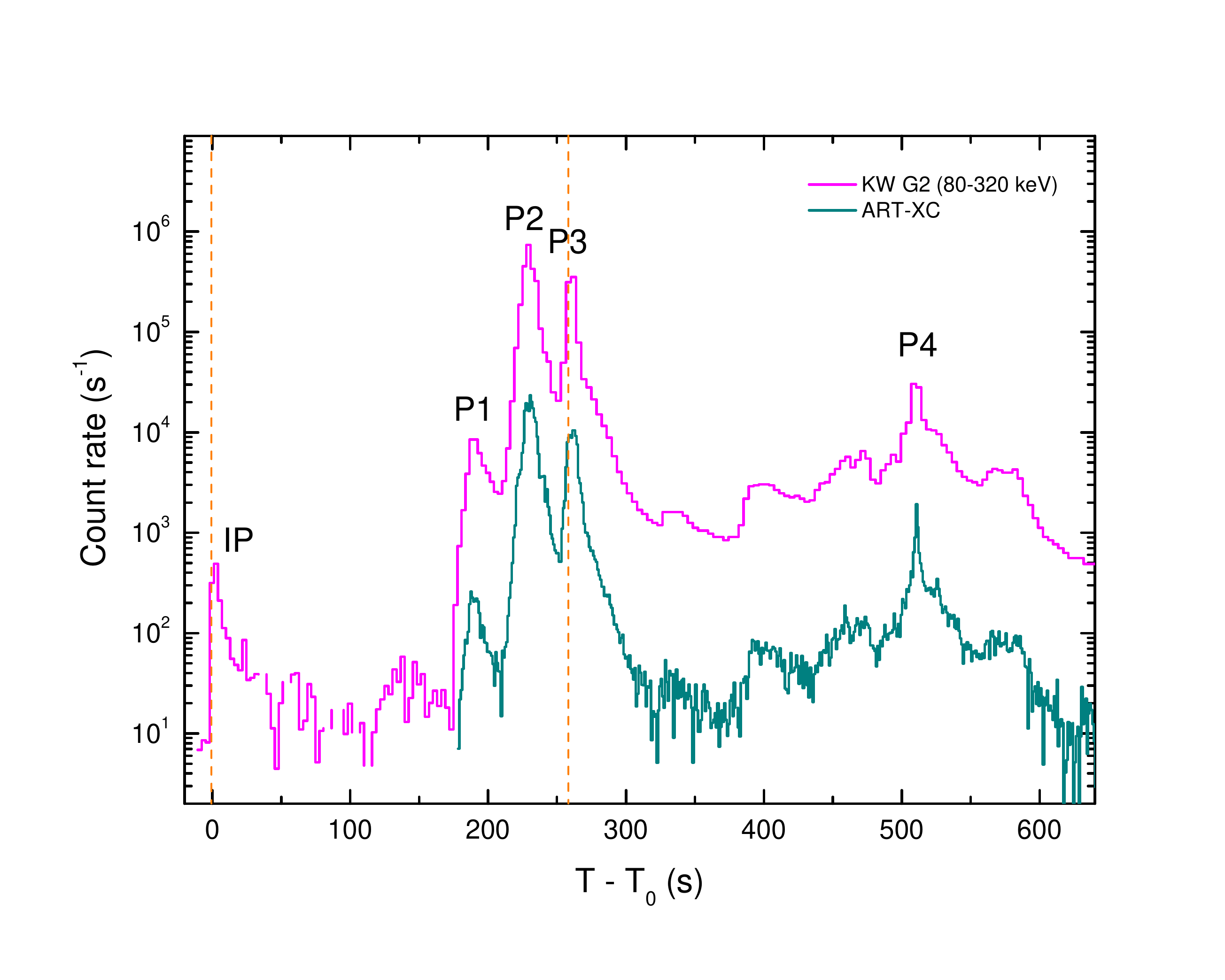}
	\caption{Overview of \GRB\ prompt emission as observed by KW and ART-XC. 
		The KW background-subtracted light curve, corrected for instrumental effects, is composed 
		of THA, BGA, and HGA count rates in G2 (80-320 keV, the magenta line).
		The dark green line shows background-subtracted ART-XC light curve in the full energy range 4-120 keV with the resolution of $1$~s.
		Labels indicate the positions of five peaks discussed in the text: the initial pulse (IP) and four prominent peaks P1-P4 during the main phase. 
		The KW triggered-mode data are available for the interval between two vertical dashed lines. 
	}
	\label{FigOverview}
\end{figure*}

\GRB\ triggered KW at $T_0(KW)$=47821.648 s UT (13:17:01.648) on 2022 October 22.
The KW trigger time corresponds to the Earth-crossing time $T_0$=47820.401~s UT (13:17:00.401) 
that is $\sim$0.4~s after the GBM trigger and $\sim$3200~s before the BAT trigger on Swift~J1913.1+1946.
Throughout the paper, we report all times with respect to this reference point unless otherwise specified.

KW \citep{Aptekar_1995SSRv...71..265A} consists of two cylindrical NaI(Tl) detectors, S1 and S2, 
mounted on the opposite sides of the the rotationally stabilized \textit{Wind} spacecraft.
The burst triggered S2, with the incident angle of $48\fdg2$ and an effective area of 90-150~cm$^2$, depending on the photon energy. 
S1 observed the burst through the spacecraft body and the rear structure of the detector, with the incident angle of $132\fdg8$. 
The attenuation of the burst emission detected in S1 cannot be easily quantified, but unsaturated data from this detector can be used as a reference.

In the triggered detector S2, count rates were recorded by time history analyzers (THA) in three energy bands:
 G1(20--80 keV), G2(80--320 keV), and G3(320--1220 keV). The record starts at $-0.512$~s; 
where 2-ms resolution light curves are available up to 0.512~s, 16-ms -- up to 33.280~s, 64-ms -- up to 98.816~s, and 256-ms -- up to 229.632~s.
Starting from $T_0$, 64 multichannel energy spectra were measured by two pulse-height analyzers: 
PHA1 (63 channels, 20-1200~keV) and PHA2 (60 channels, 0.4-16.5~MeV). 
For spectra 1 to 56, the accumulation time varies between 64~ms and 8.192~s. 
For the last eight spectra, measured from 192.256~s to 257.792~s, the accumulation time is fixed at 8.192~s. 

The KW ``waiting mode THA'' (BGA) data are available in G1, G2, and G3 from both detectors up to $251.371$~s, 
along with the count rates at energies $\sim 16.5$--22~MeV (the Z channel), all with a resolution of 2.944~s. 
In the interval from $251.371$~s, when the measurements were stopped due to the data readout, to $\sim 5070$~s, 
when the waiting mode resumed, only the count rates in G2 are available from S2, with the time resolution 
of 3.68~s and the very rough quantization of 256 counts per time bin (the ``housekeeping THA'', hereafter HGA, data).

The reduction of the light-curve and spectral data is made using standard KW analysis tools and procedures (described, e.g., in \citealt{Svinkin_2016ApJS..224...10S,Tsvetkova_2017ApJ...850..161T,Tsvetkova_2021ApJ...908...83T,Lysenko_2022ApJS..262...32L}).
Standard KW dead-time (DT) corrections \citep{Mazets_1999AstL...25..635M} are applied to the time history and spectral data outside 
the region of the most intense emission, from 216~s to 270~s, 
where additional flux saturation and pulse-pileup corrections are required (see Appendix~\ref{app:corr} for the details).

For the prompt emission light curves (up to 650~s), we assume a constant background estimated from a preburst interval from -2500 to -150~s,
during which count rates in all energy ranges of both KW detectors are consistent with being Poisson distributed.
To analyze the burst extended emission on time-scales of tens of ks we use linear background approximations, 
constructed, for each channel, from time-averaged count rates in two time intervals: 10~ks preceding $T_0$ and from 30 to 40~ks after the trigger.
For G2 and G3 these approximations are consistent with the constant background we use for the prompt emission analysis;
however, for G1, a slight negative slope is required (mainly due to the activity of bright Galactic X-ray sources).

Background spectra were extracted during a ``quiet'' time interval from 74~s to 123~s, and the detector energy scale was calibrated using the background spectra. 
The spectral analysis is performed with \textsc{XSPEC}\footnote{\url{https://heasarc.gsfc.nasa.gov/xanadu/xspec/}} \citep{Arnaud_1996ASPC..101...17A}, 
version 12.11.1, using the $\chi^2$ statistic and the Band GRB function \citep{Band_1993ApJ...413..281B} if not mentioned otherwise:
 $f(E) \propto E^{\alpha}\exp(-(2+\alpha)E/E_{\mathrm{peak}})$ for $E < E_{\mathrm{peak}}(\alpha-\beta)/(2+\alpha)$, 
and $f(E) \propto E^{\beta}$ for $E \geq E_{\mathrm{peak}}(\alpha-\beta)/(2+\alpha)$, 
where  $\alpha$ is the low-energy photon index, $E_{\mathrm{p}}$ is the peak energy in the $\nu$F$_\nu$ spectrum, 
and $\beta$ is the photon index at higher energies. The spectral model is normalized to the energy flux in the 20~keV--10~MeV range, a standard band for the KW GRB spectral analysis.

Spectral lags $\tau_{\mathrm{lag}}$ between the KW light curves are calculated with the method similar to that used in \cite{Frederiks_2013ApJ...779..151F}.

\subsection{ART-XC} \label{sec:obsART}

ART-XC is a grazing-incidence-focusing X-ray telescope on board the \textit{SRG} observatory \citep{sunyaev21}.
The telescope includes seven independent modules and has an FoV of 36~arcmin in angular diameter.
It provides imaging, timing, and spectroscopy in the 4-30~keV energy range with the total effective area of $\sim450$\,cm$^2$ at 6~keV, 
angular resolution of $45$\arcsec, energy resolution of $1.4$ keV at 6 keV and timing resolution 
of 23 $\mu$s \citep{2021A&A...650A..42P}. The primary purpose of ART-XC is to carry out the all-sky survey in hard X-rays with unprecedented sensitivity. 
At the same time, due to the high sensitivity and wide working energy range of the detectors (4-120 keV), ART-XC is able to detect high-energy events, 
such as solar flares or GRBs, from any direction in the sky (see, e.g., \citealt{2021GCN.30283....1L}).\footnote{\url{https://monitor.srg.cosmos.ru/}} 

The instrument detected \GRB\ at 13:19:55 UT on 2022 October 9. The burst happened outside its FoV, 
but its emission is well registered with all seven detectors. 
Due to the strong attenuation of the signal passed through the surrounding matter, 
ART-XC registers a light-curve shape that is practically not distorted by instrumental effects such as pulse pileup or flux saturation.

The telescope structure is designed in such a way that X-rays from celestial sources as well as cosmic background radiation 
are completely absorbed if coming not from the FoV. However, \GRB\, came from about 30 degrees off axis 
through the lateral surface of the structure of the instrument. 
This means, that at least in the 4--60~keV energy range it did not detect the direct radiation from the burst, 
but rather saw high-energy photons, whose energies were converted in the surrounding telescope structure by means of Compton scattering. 
Therefore, in the following analysis we use all photons registered by ART-XC in the energy range of 4--120~keV 
and correct count rates on DT and efficiency of CdTl detectors. 
The data from each module are analyzed separately, and then the results are combined.

\section{Analysis and Results} \label{sec:analysis}

Figure~\ref{FigOverview} shows time history of \GRB\ reconstructed from KW and ART-XC observations. 
The burst prompt emission has a complex time profile consisting of two distinct emission episodes. 
It starts with a single initial pulse (IP), which is followed, after a period of quiescence, 
by an extremely bright emission complex that lasts for $\sim 450$~s and shows four prominent peaks: 
P1, at the onset; two huge pulses P2 and P3; and a much longer but less intense P4. 
After $\sim$600~s, the pulsed prompt phase of the burst evolves to a steadily decaying, 
extended emission tail, which is visible in the KW data for more than 25~ks.
Results of the KW spectral analysis are summarized in Table~\ref{TableSpec}.

\subsection{Initial Pulse (IP)} \label{sec:IP}

The light curve of the smooth, FRED-like IP, which triggered KW, resembles that of a typical long GRB. 
It starts at -1.8~s, peaks at $\sim 0.8$~s and decays to $\sim 30$~s, with the G2 durations
$T_\mathrm{90}$ and $T_\mathrm{50}$ of $(29.9 \pm 3.9)$~s and $(10.4 \pm 1.0)$~s, 
respectively\footnote{$T_\mathrm{90}$ and $T_\mathrm{50}$ are the times to detect 90\% and 50\% of the observed count fluence, respectively.}.
The peak count rate is reached at $1.10\times10^{3}$~counts~s$^{-1}$ in the 64~ms interval starting from 0.768~s. 

A time-averaged spectrum of the IP, measured from $T_0$ to 24.832~s, is best described by 
an exponentially cut off power-law (CPL), parameterized as $E_p$: 
$f(E) \propto E^{\alpha} \exp (-E(2+\alpha)/E_p)$, with $\alpha\approx-1.62$ and $E_p \approx 970$~keV. 
A CPL fit to the spectrum near the peak count rate (from $T_0$ to 8.448~s) is characterized 
by a similar $\alpha\approx-1.65$ and the considerably higher $E_p \approx 1500$~keV. 
For both spectra, fits with the Band function are poorly constrained and set only 
an upper limit on the high-energy photon index ($\beta <-2.0$). 
The energy fluence of the IP is $(2.56 \pm 0.52)\times10^{-5}$~erg~cm$^{-2}$ 
and the 64~ms peak energy flux is $(6.20 \pm 1.52)\times10^{-5}$~erg~cm$^{-2}$~s$^{-1}$ (both in the 20~keV--10~MeV energy range).

For the IP, we derived statistically significant spectral lags ($\tau_\mathrm{lag}$) between 
the 64-ms light curves in G3 and G2 ($280\pm97$~ms), and between G3 and G1 light curves ($180\pm86$~ms). 
The positive spectral lags are indicative of hard-to-soft spectral evolution.

\begin{deluxetable*}{clcccccc}
	\tablewidth{0pt}
	\tablecaption{KW spectral fits to the prompt emission spectra with the Band function
		\label{TableSpec}}
	\tablehead{
		\colhead{Spectrum} &\colhead{Time interval} & \colhead{$\alpha$} & \colhead{$\beta$} & \colhead{$E_{\mathrm{peak}}$} & \colhead{$\chi^2/$dof} & \colhead{Flux\tablenotemark{a,b}} \\ 
		\colhead{} &\colhead{(s)} & \colhead{} & \colhead{} & \colhead{(keV)} & \colhead{} & \colhead{(erg~cm$^{-2}$~s$^{-1})$}  
	}
	\startdata
	\\
	\multicolumn{7}{c}{Initial pulse} \\[0cm]
	\\
	1 to 7		& 0.000--24.832		& -1.62$_{-0.04}^{+0.05}$ & $<2.0$ & 970$_{-330}^{+704}$ & $80/98$  & 1.03$_{-0.15}^{+0.21}\times10^{-6}$ \\
	1 to 5		& 0.000--8.448 		&-1.65$_{-0.03}^{+0.03}$ & $<2.0$ & 1495$_{-448}^{+884}$ & $89/94$  & 2.58$_{-0.30}^{+0.40}\times10^{-6}$ \\
	\\
	\multicolumn{7}{c}{P1} \\[0cm]
	\\
	27 to 58	& 180.480--208.640 	&-1.17$_{-0.02}^{+0.02}$ & -2.60$_{-0.12}^{+0.09}$ & 1011$_{-54}^{+54}$ & $181/97$  & 3.64$_{-0.04}^{+0.04}\times10^{-5}$\\
	35 to 41	& 186.624--188.416 & -0.93$_{-0.02}^{+0.02}$ & -3.08$_{-0.48}^{+0.27}$ & 1702$_{-110}^{+111}$ & $58/79$  & 1.11$_{-0.05}^{+0.05}\times10^{-4}$ \\
	\\
	\multicolumn{7}{c}{P2} \\[0cm]
	\\
	59			& 208.640--216.832 	&-1.33$_{-0.03}^{+0.03}$ & -2.40$_{-0.18}^{+0.12}$ & 981$_{-106}^{+112}$ & $158/93$  & 3.67$_{-0.09}^{+0.09}\times10^{-5}$\\
	60			& 216.832--225.024 	&-1.18$_{-0.07}^{+0.08}$ & -2.49$_{-0.06}^{+0.05}$ & 2733$_{-133}^{+141}$ & $60/55$  & 1.29$_{-0.18}^{+0.20}\times10^{-3}$\\
	61\tablenotemark{c}	& 225.024--233.216 	&-0.76$_{-0.05}^{+0.05}$ & -2.13$_{-0.02}^{+0.02}$ & 3038$_{-116}^{+120}$ & $75/59$  & 1.62$_{-0.09}^{+0.09}\times10^{-2}$\\
	62			& 233.216--241.408 	&-0.86$_{-0.06}^{+0.06}$ & -2.78$_{-0.04}^{+0.04}$ & 1617$_{-32}^{+32}$ & $77/58$  & 2.17$_{-0.09}^{+0.09}\times10^{-3}$\\
	63			& 241.408--249.600 	&-1.25$_{-0.04}^{+0.04}$ & -2.75$_{-0.06}^{+0.05}$ & 1072$_{-31}^{+32}$ & $85/67$  & 3.15$_{-0.15}^{+0.15}\times10^{-4}$\\
	\\
	\multicolumn{7}{c}{P3} \\[0cm]
	\\
	64			& 249.600--257.792 	&-0.97$_{-0.06}^{+0.07}$ & -2.51$_{-0.04}^{+0.04}$ & 1886$_{-71}^{+73}$ & $53/56$  & 8.60$_{-0.12}^{+0.12}\times10^{-4}$\\
	\\
	\multicolumn{7}{c}{ P1 + P2 + P3} \\[0cm]
	\\
	27 to 64\tablenotemark{d}    & 180.480--257.792 	&-0.89$_{-0.05}^{+0.06}$ & -2.21$_{-0.02}^{+0.02}$ & 2660$_{-105}^{+109}$ & $60/58$  & 2.22$_{-0.10}^{+0.11}\times10^{-3}$\\
 	\\
	\enddata
	\tablenotetext{a}{\footnotesize{Averaged over the spectrum accumulation time interval.}}
	\tablenotetext{b}{\footnotesize{In the 20~keV--10~MeV band.}}
	\tablenotetext{c}{\footnotesize{``Peak'' spectrum, used to calculate the peak energy flux.}}
	\tablenotetext{d}{\footnotesize{Time-averaged spectrum, used in calculation of the prompt emission fluence.}}
	\end{deluxetable*}

\subsection{The Main Phase} \label{sec:MP}

\begin{figure*}
	\centering
	\includegraphics[width=0.7\textwidth]{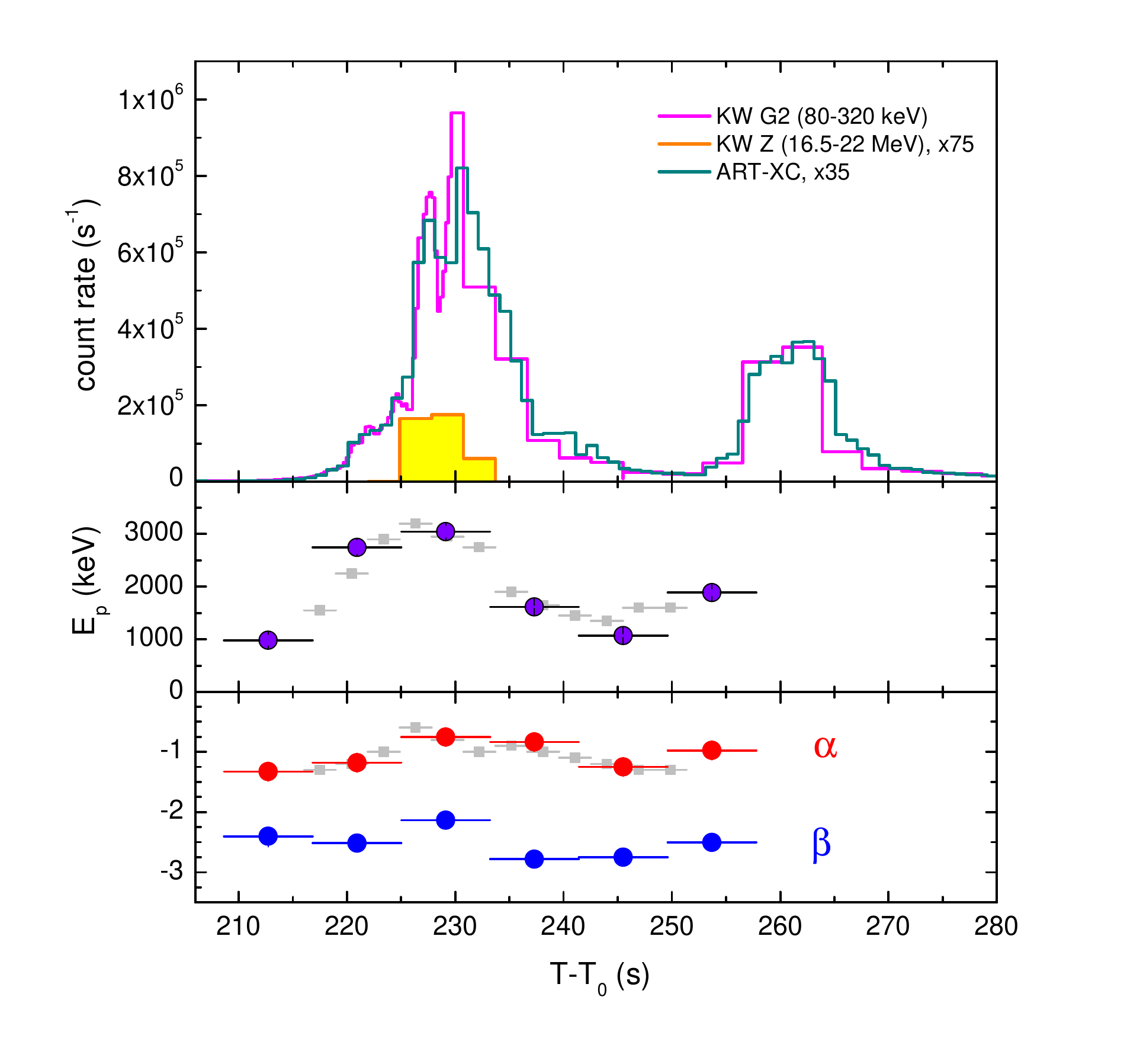}
	\caption{Brightest phase of \GRB\ (peaks P2 and P3). 
		The upper panel shows the light curve as seen by KW in G2 (80--320~keV, dead-time and pileup corrected count rate, magenta), ART-XC (dead-time corrected count rate times 35, dark green), and by KW in the Z band (16.5--22~MeV, dead-time corrected count rate times 75, orange/yellow). Middle panel: temporal evolution of the spectral peak energy $E_p$ as derived from the KW spectral fits with the Band function (Table~\ref{TableSpec}). Lower panel: the evolution of the model photon indices: low-energy $\alpha$ (red) and high-energy $\beta$ (blue). For the spectral parameters, statistical errors are within the data points. Gray points illustrate $E_p$ and $\alpha$ estimates obtained from the KW light-curve data (Appendix~\ref{app:corrLC}).  
	}
	\label{FigP2P3}
\end{figure*}

During about 150~s after the end of the IP, the emission barely exceeds the background level, 
with only a hint of a wide bump around $\sim150$~s in the KW light curve. 
The main phase of the event begins at 175~s with a fast rise of the emission intensity 
to the peak P1 around 188~s (with the peak count rate of $\sim1\times10^{4}$~counts~s$^{-1}$, or 10x the IP),
then temporary decays to $\sim 2 \times10^{3}$~counts~s$^{-1}$ around 208~s, the minimum between P1 and P2.
A time-averaged spectrum of this pulse, measured from 180.48~s to 208.64~s is best described by a Band function 
with $\alpha\approx-1.17$, $\beta\approx-2.60$ and $E_p \approx 1010$~keV.

The brightest phase of the burst (from $\sim208$~s to $\sim280$~s) is shown in Figure~\ref{FigP2P3}. 
Two huge pulses P2 and P3 contain about $\sim90$\% ($\sim60$\% and $\sim30$\%, respectively) 
of the total burst counts recorded in both KW and ART-XC light curves.
In the KW G2 band (and in the combined G1+G2+G3 band 20--1220~keV),
the enormous peak count rate of $\sim9.6\times10^{5}$~counts~s$^{-1}$ ($\sim2.2 \times10^{6}$~counts~s$^{-1}$) is reached 
in a 1~s interval around $T_\mathrm{peak}=230$~s, at the second peak of the double-peaked P2. 
The ART-XC light curve shows a similar pattern, with the 1~s peak count rate of $\sim2.7\times10^{4}$~counts~s$^{-1}$.
The emission at this phase is not only extremely intense but also spectrally hard: for the first time in KW GRB observations, 
a statistically significant ($>60$~$\sigma$) excess over the background is detected in the instrument's Z channel ($\sim 16.5$--22~MeV), 
which lasts for about 10~s and peaks, at $\sim 2400$ counts/s, at the same time as the sub-MeV emission.

\begin{figure*}
	\centering
	\includegraphics[width=0.59\textwidth,height=0.49\textwidth]{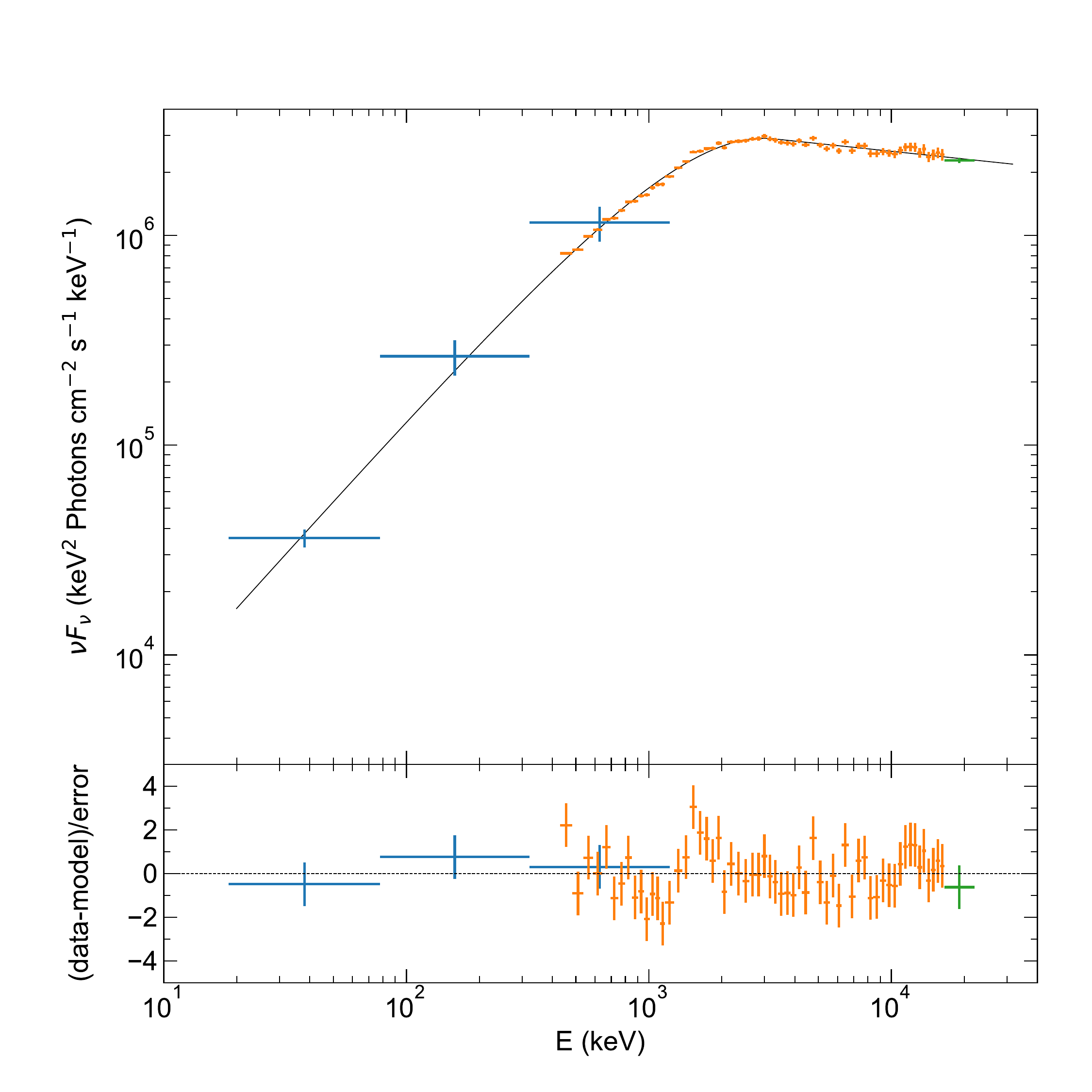}
	\caption{$\nu F_\nu$ spectrum at the peak of the prompt emission (225~s to 233~s). Blue points represent pileup- and saturation-corrected THA data; orange points: pileup- and saturation-corrected PHA2 data; and the green point - dead-time corrected Z-channel data (16.5--22~MeV). The best spectral fit with the Band function (Table~\ref{TableSpec}) is shown with the solid line.  
	}
	\label{FigSpec}
\end{figure*}

Spectral fits during the brightest phase are made using the pileup- and saturation-corrected THA (20--1220~keV) and PHA2 (0.4--16.5~MeV) data, 
and, at the highest peak, the DT-corrected Z-channel data (Figure~\ref{FigSpec}). 
PHA1 data cannot be used due to the unrecoverable instrumental effects arising at such enormous fluxes (Appendix~\ref{app:corrPHA}).
The two lower panels in Figure~\ref{FigP2P3} show the spectral evolution of the emission: $E_p$ rises from $\sim$1~MeV between P1 and P2 
to $\sim$3~MeV around the peak count rate, then drops back to $\sim$1~MeV between P2 and P3, and rises again to $\sim$2 MeV during the first $\sim$1/3 of the second huge pulse P3 
(the last spectrum measured by KW). The temporal evolution of the low-energy photon index $\alpha$ shows a similar pattern,
which is consistent with a positive correlation between the emission intensity and its spectral hardness.

After the peak of P3 around 260~s the burst intensity starts to drop drastically (nearly to the pre-event level in ART-XC) but then increases again. The final, less bright phase of the prompt emission (P4) has a long ($\sim 300$~s), complicated structure, with the narrow count-rate maximum around 510~s. This part of the event was observed by KW in a single energy band (G2), making its spectral analysis impossible. 

For the main phase, and given the relative weakness of IP, for the whole prompt emission, durations $T_\mathrm{90}$ and $T_\mathrm{50}$ in the KW 80-320~keV band are $284.0 \pm 3.7$~s and $31.1 \pm 3.7$~s, respectively.
Estimated from the ART-XC ligt curve, the durations are very similar, $T_\mathrm{90}=276.0 \pm 5.4$~s and $T_\mathrm{50}=31.0 \pm 1.4$~s.

\subsection{Observer-frame Energetics in the Prompt Emission}\label{sec:fluence}

A time-averaged spectrum of the main phase of the prompt emission (180~s to 258~s, Table~\ref{TableSpec} )
is best described by a Band function with $\alpha\approx-0.89$, $\beta\approx-2.21$ and $E_p \approx 2660$~keV.
From this spectrum, the energy fluence measured up to the end of the KW triggered mode is $(0.172\pm 0.015)$~erg~cm$^{-2}$. 
Using the KW count-to-fluence ratio for the last recorded spectrum and assuming that the emission hardness during the remaining part of P3 is not much different, 
we calculate the overall fluence in P1+P2+P3 to be $(0.21\pm 0.017)$~erg~cm$^{-2}$.

The lack of KW spectral data for P4 does not allow us evaluate its fluence directly. 
Therefore, using the fraction of the total KW counts in this pulse ($\sim10$\%), 
and under the assumption that emission at this stage is likely softer than in the huge peaks (e.g., \citealt{Kann_2022GCN.32762....1K}), 
we account for the P4 contribution by adding 5\% ($\approx0.01$~erg~cm$^{-2}$) to the P1+P2+P3 fluence and 2.5\% systematic to the uncertainty. 
As a result, we obtain the total energy fluence of the prompt emission $S=(0.22 \pm 0.02)$~erg~cm$^{-2}$ 
(0 to 600~s, 20~keV--10~MeV). %The time-averaged spectrum reported above covers about 80\% of the fluence. 

The spectrum at the brightest emission peak (225.024 -- 233.216~s) is best fit with $\alpha\approx-0.76$, 
$\beta\approx-2.13$ and $E_p \approx 3040$~keV. 
From this spectrum and a peak-to-average count-rate ratio in the combined G1+G2+G3 
light curve\footnote{Calculations using the KW spectrum and the ART-XC light curve 
yield very similar $F_{\mathrm{p}}$ value.} 
we calculate the 20~keV--10~MeV peak energy flux of the burst $F_{\mathrm{p}}=(3.14\pm0.47)\times 10^{-2}$~erg~cm$^{-2}$~s$^{-1}$ (or $\sim 1.4\times 10^{4}$~photons~cm$^{-2}$~s$^{-1}$), in a 1-s interval starting from 229.632~s. 
As for the exceptionally high count rate, the derived $F_{\mathrm{p}}$ is the highest among $>$3500 GRBs detected by the KW so far.

\subsection{Early Afterglow} \label{sec:ext}

\begin{figure*}
	\centering
	\includegraphics[width=0.8\textwidth]{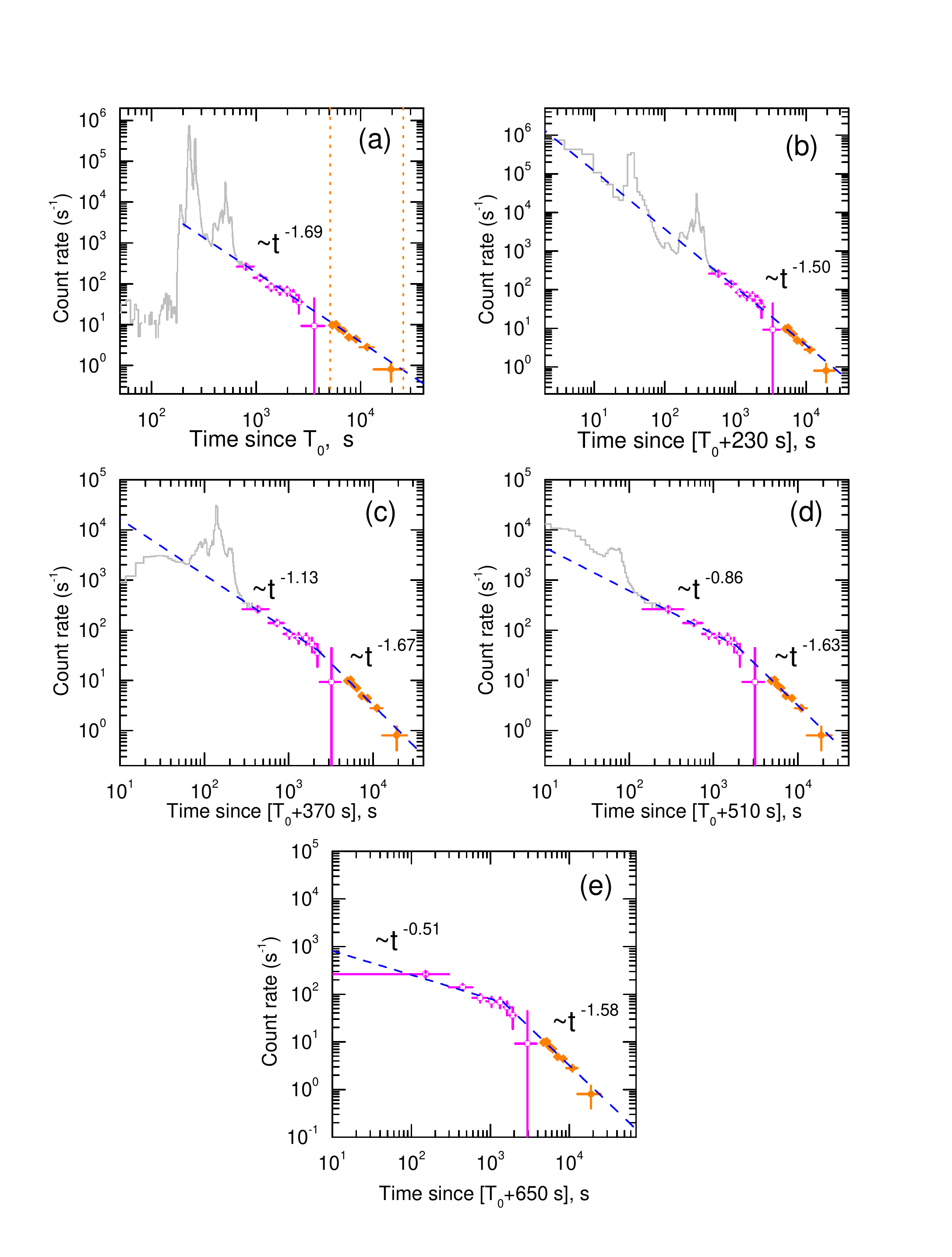}
	\caption{Early afterglow of \GRB\ observed by KW in the 80--320~keV band. 
		Magenta points: time-averaged HGA data. Orange points: time-averaged BGA data. 
		The interval from 5.1~ks to 25.7~ks (the vertical dotted lines in panel (a)) is used for the late-time spectral fit.
		Dashed lines in each panel show the best PL (or the best BPL) temporal fit to the data in the interval from 650~s to 25.7~ks after the trigger.
		Zero time points of the fits are specified in the x-axis labels. The prompt emission light curve (gray solid lines) is shown for the reference. 
	}
	\label{FigAfter}
\end{figure*}

The pulsed prompt phase of the burst ends at $\sim600$~s, when the light curve evolves to a steadily decaying emission tail, 
which is below the sensitivity of ART-XC but is visible in the KW data for more than 25~ks (Figure~\ref{FigAfter}a).
In the KW G2 band, the decay in the interval from $650$~s to $25.7$~ks is well described by a simple power law (PL) $N(t)\propto (t-T_0)^{-\alpha_t}$ 
with the PL index $\alpha_\mathrm{t} = 1.69\pm0.03$ ($\chi^2$=16/14 dof), while a broken PL (BPL ) fit to the data is not constrained.

Starting from 5.1~ks, count rates in all three bands (G1, G2, and G3) are available that allows estimating emission spectrum.
From a PL fit to a three-channel spectrum constructed for the time interval from 5.1~ks to 25.7~ks,
we obtain the photon index $\Gamma=1.99\pm0.05$ and time-averaged flux $1.10_{-0.08}^{-0.06}\times 10^{-8}$~erg~cm$^{-2}$~s$^{-1}$ (20--1500~keV). 
Assuming a PL spectrum with $\Gamma$=2, we estimate the 20~keV--10~MeV energy fluence of the \GRB\, 
extended emission from 650~s to 25.7~ks to be $(2.15\pm0.14)\times 10^{-3}$~erg~cm$^{-2}$,
or $\approx1$\% of the energy in the prompt phase of the burst. 
Using the late-time spectrum, we extrapolate KW flux points after $\sim 5$~ks to the 0.3--10~keV band and find them consistent, 
within a factor of $\sim$1.3, with unabsorbed fluxes derived from simultaneous XRT observations\footnote{Unabsorbed \textit{Swift}-XRT fluxes were extracted 
from the XRT repository 
\citep{Evans_2007A&A...469..379E,Evans_2009MNRAS.397.1177E}.}. 

The combination of the spectral and temporal behaviors of the steadily decaying emission is in reasonable agreement with that expected 
at the ``normal''(III) phase of the canonical X-ray afterglow \citep{Nousek_2006ApJ...642..389N,Zhang_2006ApJ...642..354Z,Racusin_2009ApJ...698...43R}
and supports a scenario in which the bright, extended $\gamma$-ray emission observed by KW 
is generated by the synchrotron forward-shock mechanism during the normal spherical decay of the afterglow \citep{Meszaros_Rees_1997ApJ...476..232M}.
It should be noted, however, that the use of $T_0$, corresponding to the early and relatively weak precursor, 
as a zero time point ($t_0$) of the bright afterglow can barely be justified, 
and the decay slope measured at times not much larger than $T_\mathrm{90}$ could  
be very sensitive to the assumed $t_0$ (the ``$t_0$ effect''; \citealt{Zhang_2006ApJ...642..354Z}).

Therefore, in order to characterize the afterglow temporal behavior more precisely and to identify a possible break, 
we performed temporal PL and BPL fits with $t_0$ set to several characteristic times in the \GRB\ light curve: 
$T_\mathrm{peak}=230$~s, the peak time of the prompt emission; 370~s, the light-curve minimum between the brightest phase (P2+P3) 
and the last episode of the prompt emission (P4); 510~s, the peak time of P4; 
and, finally, 650~s, the time when the steadily decaying afterglow starts to dominate the observed flux.

The results of our fits with different zero time points are presented in Table~\ref{TableFitsAfter} and Figure~\ref{FigAfter}.
The choice of $T_\mathrm{peak}$ as $t_0$ does not constrain a break and results in a more gentle, as compared to $t_0=0$, PL slope $\alpha_\mathrm{t} = -1.50\pm0.04$. 
This index is in perfect agreement with the soft X-ray slope of $\approx 1.5$ between $\sim3$~ks and $\sim80$~ks \citep{OConnor_2023arXiv230207906O,Williams_2023ApJ...946L..24W}, 
and it also fits better in the slope range of the ``normal'' afterglow segment ($1<\alpha_t<1.5$). 

With the shift of $t_0$ toward the end of the prompt emission, a broken PL shape becomes preferred by the data: 
the break significance increases from $\sim2.5\sigma$ ($t_0=370$~s) to $\sim4\sigma$ ($t_0=510$~s) and $\sim10.5\sigma$ ($t_0=650$~s).
In the two latter cases, the combination of a shallow pre-break slope $\alpha_\mathrm{1,t}\sim 0.51-0.86$ 
and the steeper post-break slope $\alpha_\mathrm{2,t}\sim1.6$ closely resembles that of the transition from 
the ``plateau'' (Segment II) to the ``normal'' phase of the canonical X-ray afterglow. 
The break positions, located in a narrow time interval of the light curve (2100--2600~s relative to $T_0$), 
are also in the range expected for a break from Segment~II to Segment~III ($10^3-10^4$~s).
We note, that, using $\sim T_0+510$~s as a reference time point when fitting the late-time GBM light curve, \cite{Lesage_2023arXiv230314172L} 
obtained a similar index to KW PL index ($\sim 0.82$) for the decay from 650~s to 1460~s after the trigger (after that time \GRB\ is occulted by Earth for \textit{Fermi}); 
and also the post-break KW index is consistent with the early soft X-ray slope of $\approx1.5$ noticed above in this Section.  

%Finally, the absence of a shallow plateau segment in the case 

\begin{deluxetable*}{ccccccccc}
	\tablewidth{0pt}
	\tablecaption{KW temporal power-law fits to the early afterglow
		\label{TableFitsAfter}}
	\tablehead{
		\colhead{$t_0$} & \colhead{Model} & \colhead{$\alpha_t$} &  \colhead{$\alpha_\mathrm{1,t}$} & \colhead{$\alpha_\mathrm{2,t}$} & \colhead{$t_b$}                & \colhead{$\chi^2$/dof} \\ 
		\colhead{} 				   & \colhead{}      & \colhead{}           & \colhead{}                       & \colhead{}                      & \colhead{(s)}       & \colhead{}   
	}
	\startdata
	 0 			               & PL              & $1.69_{-0.03}^{+0.03}$ & ...                        & ...                             & ...                            & 16/14 \\
	 \\
	 230~s 	               & PL              & $1.50_{-0.04}^{+0.04}$ & ...                            & ...                             & ...                            & 14/14 \\
%	 230~s 	               & BPL             &                         & $1.35_{-0.12}^{+0.14}$        & $1.71_{-0.10}^{+0.11}$           & $2730_{-1030}^{+720}$         & 10/12  \\
	 \\
	 370~s 	               & PL              & $1.36_{-0.02}^{+0.02}$ & ...                            & ...                              & ...                           & 23/14 \\
	 370~s 	               & BPL             & ...                    & $1.12_{-0.10}^{+0.10}$        & $1.67_{-0.09}^{+0.12}$           & $2200_{-540}^{+870}$          & 9/12  \\
	 \\
	 510~s 	               & PL              & $1.20_{-0.02}^{+0.02}$ & ...                           & ...                              & ...                           & 31/14  \\
	 510~s 	               & BPL             & ...                    & $0.86_{-0.08}^{+0.08}$        & $1.63_{-0.09}^{+0.11}$           & $1840_{-370}^{+550}$          & 8/12  \\
	 \\
	 650~s 	               & PL              & $0.88_{-0.01}^{+0.01}$ & ...                           & ...                               & ...                           & 141/14\\
	 650~s 	               & BPL             & ...                    & $0.51_{-0.04}^{+0.04}$       & $1.58_{-0.09}^{+0.10}$            & $1440_{-250}^{+300}$          & 9/12 \\
	\\
	\enddata
\tablecomments{\footnotesize{The fits are made in the time interval from 650~s to 25.7~ks and use $t_0$ as a zero time point. The model parameters are as follows: $\alpha_t$ is the simple PL index; $\alpha_\mathrm{1,t}$ and $\alpha_\mathrm{2,t}$ are BPL pre-break and post-break indices, respectively; and $t_b$ is the break time (with respect to $t_0$).}}
\end{deluxetable*}

\begin{figure*}
	\begin{center}
		\includegraphics[width=0.89\textwidth,angle=0]{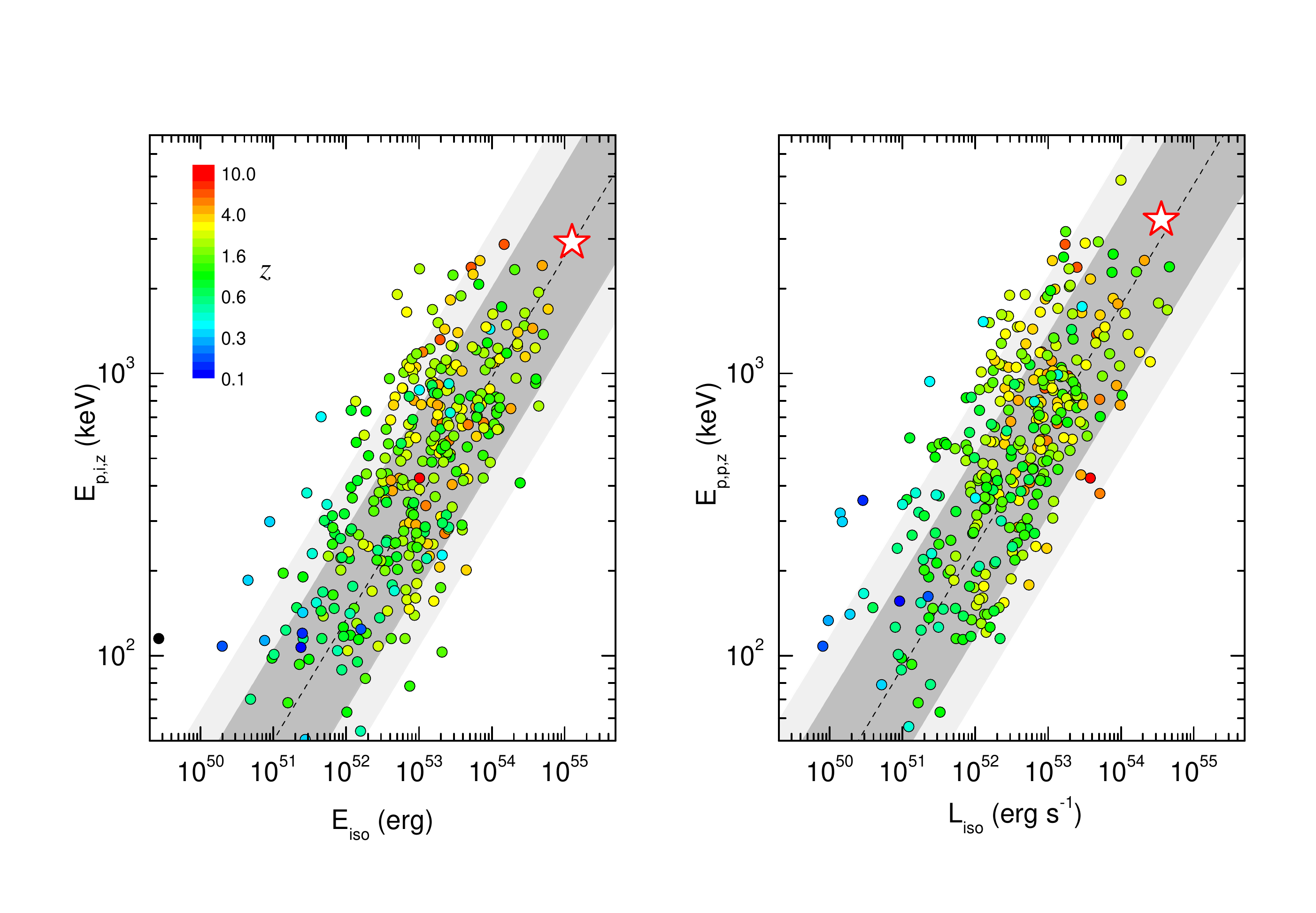}
		\caption{Rest-frame energetics of \GRB\ in the $E_\mathrm{p,i,z}$--$E_\mathrm{iso}$ and $E_\mathrm{p,p,z}$--$L_\mathrm{iso}$ planes (stars).
			The rest-frame parameters of 315 long KW GRBs with known redshifts \citep{Tsvetkova_2021ApJ...908...83T} are shown with circles; 
			the color of each data point represents the burst's redshift.
			The `Amati' and `Yonetoku' relations for this sample are plotted with dashed lines and the dark- and light-grey shaded areas show their 68\% and 90\% prediction intervals, respectively. The error bars are not shown here for reasons of clarity. 
		}
		\label{fig:amyo}%
	\end{center}
\end{figure*}

\section{Discussion}\label{sec:discussion}
\subsection{Prompt Emission in Context}

\GRB\ is the brightest gamma-ray burst observed by KW in almost 30 years of observations.
The incident photon flux, peaking at $\gtrsim 2 \times 10^6$~s$^{-1}$, 
was previously exceeded only in KW detections of giant flares from 
Galactic magnetars SGR~1900+14 (on 1998 August 27; \citealt{Mazets_1999AstL...25..635M})
and SGR~1806-20 (2004 December 27; \citealt{Frederiks_2007AstL...33....1F}), 
and is comparable to that in the extremely bright flare from SGR~1627-41 (1998 June 18; \citealt{Mazets_1999ApJ...519L.151M}). 

Since the launch in 1994 November and up to 2023 February, KW has detected $\sim 3570$ GRBs 
with virtually no bright GRBs having been missed. About $\sim 85$\% of them are long-duration bursts. 
To compare \GRB\ with the bright KW GRB population, we have selected $\sim 150$ long bursts with $S\gtrsim 10^{-4}$~erg~cm$^{-2}$. 
In this sample, GRB\,130427A has the largest fluence $3 \times 10^{-3}$~erg~cm$^{-2}$ and GRB\,140219A 
is the brightest in terms of the peak energy flux ($1.2 \times 10^{-3}$~erg~cm$^{-2}$~s$^{-1}$). 
Ten percent of the most fluent bursts have durations ranging from 10~s (GRB\,021206) to 680~s (GRB\,160625B) and $E_p$ of time-integrated spectra between $\sim 550$~keV (GRB\,160625B) and $\sim 2500$~keV (GRB\,140219A). 
The KW ultra-long GRBs (22 bursts with a duration $\gtrsim 1000$~s) have moderate spectral hardness, with typical $E_p$ of a few hundred keV, which yield moderate fluences below $6 \times 10^{-4}$~erg~cm$^{-2}$.
Thus, \GRB\ is at the extreme end of the bright GRB population being simultaneously very long-duration and hard-spectrum. 
A more detailed analysis of \GRB\ in the context of the bright GRB population is presented in a separate paper \citep{Burns_2023ApJ...946L..31B}. 

\subsection{Rest-frame Energetics and Prompt Hardness-Intensity Correlations}

Using $z = 0.151$, and the values of the total energy fluence $S$ and the peak energy flux $F_p$ 
in the observer frame (Section~\ref{sec:fluence}), we estimate the rest-frame energetics 
of the burst prompt emission. 
Assuming a $\Lambda$CDM cosmological model with $\Omega_M = 0.308$, $\Omega_{\Lambda} = 0.692$, 
and $H_0 = 67.8$~km~s$^{-1}$~Mpc$^{-1}$ \citep{Planck2016a}, %($d_L=2.30\times10^{27}$~cm), 
$E_{\mathrm{iso}}$ is $(1.2\pm0.1)\times10^{55}$~erg, and the peak isotropic 
luminosity $L_{\mathrm{iso}}$ is $(2.1\pm0.4)\times10^{54}$~erg~s$^{-1}$ (on the 1~s scale). 
By applying a typical, for KW long GRBs, 1024~ms $F_p$ to 64~ms $F_p$ conversion factor of 1.71, 
the 64~ms $L_{\mathrm{iso}}$ is estimated to be $(3.4\pm0.5)\times10^{54}$~erg~s$^{-1}$. 
The reported energetics are calculated in the bolometric rest-frame range 1~keV--10~MeV.
Derived from the observer-frame $E_p$ values (Section~\ref{sec:fluence}), 
the rest-frame spectral peak energies $(1 + z)\ E_p$  are $E_\mathrm{p,i,z} \approx 2900$~keV (time-averaged) 
and $E_\mathrm{p,p,z} \approx 3500$~keV (at the peak luminosity). 

These estimates make \GRB\, the most energetic and the third-most luminous\footnote{After GRB~110918A \citep{Frederiks_2013ApJ...779..151F} and GRB~210619B \citep{Svinkin_2021GCN.30276....1S}.} $\gamma$-ray burst observed since the beginning of the cosmological era in 1997. 
Figure~\ref{fig:amyo} shows $E_{\mathrm{iso}}$ and $L_{\mathrm{iso}}$ for \GRB\,
along with the KW sample of more than 300 long GRBs with known redshifts
\citep{Tsvetkova_2017ApJ...850..161T,Tsvetkova_2021ApJ...908...83T}.
In the rest-frame hardness-intensity plane $E_\mathrm{p,i,z}-E_\mathrm{iso}$ \GRB\ lies inside 
the 68\% prediction interval of the ``Amati'' relation for the KW sample. 
Likewise, in the $E_\mathrm{p,p,z}-L_\mathrm{iso}$ plane, the burst perfectly fits the ``Yonetoku'' relation. 
From this, we conclude that \GRB\ is most likely a very rare, very hard, super-energetic version of a ``normal'' long GRB.

\subsection{Fundamental Plane Correlation between Prompt and Afterglow Emissions}

In Section~\ref{sec:ext} we show that, with the zero time point shifted close to the end of the prompt emission,
the broken power-law behavior of the bright, early $\gamma$-ray afterglow observed by KW can be interpreted 
as the transition from Segment~II (plateau phase) to Segment~III (normal spherical decay phase) 
of the canonical X-ray afterglow. 
Based on this assumption, we test the rest-frame parameters of the  prompt emission and the 
early afterglow against a 3-dimensional relation between the peak prompt luminosity $L_\mathrm{peak}$, 
the rest-frame time at the end of the X-ray plateau $T_X^*$, and its corresponding luminosity in X-rays $L_X$: 
the so-called 3D Dainotti fundamental plane relation \citep{Dainotti_2017ApJ...848...88D,Dainotti_2020ApJ...904...97D}. 
Given the most significant break position ($t_b=1440$~s; $t_0=650$~s), $T_X^*=t_b/(1+z)$ is $\sim1250$~s and $L_X=4 \pi d_L^2 F_X=4.0\times10^{49}$~erg~s$^{-1}$, 
where $F_\mathrm{X}=6.0\times10^{-7}$~erg~cm$^{-2}$~s$^{-1}$ is the X-ray flux at $t_b$, extrapolated from the KW band to the 0.3--10~keV band 
using the late-time KW spectrum\footnote{With the photon spectral index $\Gamma=2$; for this spectrum cosmological $K$-correction is unity}. 

Using $T_X^*$, $L_X$, and $L_\mathrm{peak} = L_\mathrm{iso}$ (1-s scale), we calculate a distance from \GRB\ 
to the fundamental plane for the full sample of 222 GRBs studied in \cite{Dainotti_2020ApJ...904...97D}
and to the planes for its ``gold`` (65 GRBs) and ``long GRB'' (129 events) subsamples.
In each case, we find the distance within 1 sigma scatter of the tested relation, 
with the best agreement achieved for the ``gold`` and long GRB fundamental planes. 
This further supports the consistency of \GRB\ properties with the less-energetic long GRB population.
 
\subsection{Collimation-corrected Energy and Central Engine}

Long GRBs are thought to originate in the collapse of massive stars \citep{Paczynski_1998ApJ...494L..45P,MacFadyen_1999ApJ...524..262M}.
The most widely discussed models of central engines are newborn, rapidly rotating compact objects,
such as magnetars and black holes emitting highly collimated, ultra-relativistic jets (fireballs). 
When the tightly collimated relativistic fireball is decelerated by the circumburst medium (CBM) 
down to the Lorentz factor $\approx 1/\theta_\mathrm{jet}$ ($\theta_\mathrm{jet}$ is the jet opening angle), 
an achromatic break (jet break) should appear, in the form of a sudden steepening in the GRB afterglow light curve, 
at a characteristic time $t_\mathrm{jet}$. When the opening angle of the jetted outflow is known, 
the isotropic-equivalent energetics can be converted to the more accurate collimation-corrected energetics \citep{Sari_1999ApJ...519L..17S}.

Given $E_\mathrm{iso}=1.2\times10^{55}$~erg and assuming a top-hat jet, the total collimation-corrected energy of \GRB\ is  
$E_\mathrm{K}\approx 10^{53}\ \textrm{erg}\ t_\mathrm{jet}^{3/4}\ (\eta_{\gamma}/0.2)^{-3/4}\ n_0^{1/4}$,
%$E_\mathrm{K,52} \approx 8.52\ t_\mathrm{jet}^{3/4}\ (\eta_{\gamma}/0.2)^{-3/4}\ n^{1/4}$.
where $n_0$ is the medium number density, $\eta_{\gamma}$ is the radiative efficiency of the prompt phase, and $t_\mathrm{jet}$ is measured in days.
Although extensive multiwavelength follow-up of \GRB\ did not reveal an apparent achromatic break in the afterglow light curve,
a number of $\theta_\mathrm{jet}$ estimates are reported, ranging from $0.7^\circ$ to $>10.7^\circ$ \citep{Kann_2023arXiv230206225K,OConnor_2023arXiv230207906O,Negro_2023ApJ...946L..21N,Williams_2023ApJ...946L..24W,An_2023arXiv230301203A},
which, assuming typical $\eta_{\gamma}=0.2$ and $n_0=1$~cm$^{-3}$, imply $E_\mathrm{K}$ from $\sim 5\times10^{51}$ to $\sim 5\times10^{53}$ and even higher.

The magnetar central-engine model, where the GRB is powered by a newborn, fast-rotating magnetar, 
predicts $E_\mathrm{K}$ below few~$\times 10^{52}$~erg (see, e.g., \citealt{Metzger_2011MNRAS.413.2031M}),  
while the accreting black hole models extend the limit on the GRB total energetics up to $\sim 10^{54}$ erg 
(see, e.g., \citealt{van_Putten_2017MNRAS.464.3219V}).
For \GRB\ to match the total released energy consistent with the predictions of the black hole central-engine model 
a collimation correction factor of $\gtrsim 10$ is required, corresponding to a top-hat jet half-opening angle 
constraint of $\theta_\mathrm{jet}<25^\circ$, or $t_\mathrm{jet} \lesssim 30$~days. 
The structured jet model suggested by \cite{OConnor_2023arXiv230207906O} allows even lower
total energy of the explosion, below $\sim 10^{53}$~erg, which may fit magnetar central-engine models. 
Thus, despite the enormous isotropic energy implied, an energy budget of \GRB\ can still be explained within 
standard scenario for the central engine/progenitor of long GRBs. 

A more detailed discussion of \GRB\ collimated energetics 
is presented in \cite{Burns_2023ApJ...946L..31B}, including in the context of the KW sample.

\subsection{Emission Feature around 10~MeV}

\cite{Ravasio_2023arXiv230316223E}, hereafter R23, analyzed \textit{Fermi}/GBM spectral data outside the time interval affected 
by saturation (called Bad Time Interval, BTI; 219-277~s after the GBM trigger) and discovered a highly significant 
narrow emission feature on top of the prompt emission continuum. The spectral line, modeled by a Gaussian with 
a roughly constant width $\sim1$~MeV and a central energy $E_\mathrm{line}$ decreasing in time from $\sim$12.5 to $\sim$6~MeV,
is detected at $>6\sigma$ in the interval from 280~s to 320~s (the decay phase of the second huge pulse P3) 
and nondetected before BTI (including the interval 184-216~s at the rising front of the brightest phase of the burst).   
R23 interpret this feature as a blue-shifted electron-positron annihilation line 
of relatively cold electron-positron pairs, which could have formed within the jet region 
where the brightest pulses of the GRB were produced.

Among the many $\gamma$-ray detectors that observed \GRB\, KW is one of the few instruments capable of making 
detailed spectral measurements at energies around 10~MeV and its independent identification 
of the spectral line reported from GBM data would be of obvious importance. 
Unfortunately, the time span of KW spectral measurements ends $\sim$20~s before the time range 
of the spectral line detection reported in R23. 
However, KW spectral data on the brightest part of the burst, corrected for pileups and saturations,  
can be tested for the presence of a similar spectral feature.

We visually inspect best-fit residuals for five KW spectra covering the interval 
from 217~s to 257~s, and only in one, measured at the very peak of the emission (225~s to 233~s), 
we find a marginal ($\lesssim 2\sigma$) excess in the count rate over the fitted continuum in the region around 10~MeV. 
Although this excess is not alone in this spectrum (there is another one around 1.8~MeV), 
and the systematic variations in the fit residuals may be, among other reasons, 
due to a spectral evolution of the emission during the 8-s accumulation interval, 
we analyze the spectrum for the presence of a statistically significant feature similar to that of R23. 

For this purpose, we model the excess by adding a Gaussian line (\textsc{XSPEC} model \texttt{gauss}), 
with initial $E_\mathrm{line}$=10~MeV and width fixed to 1~MeV\footnote{Fits with line width left free are not constrained.}, 
to the best-fit continuum for this spectrum.
Our fit with the combined model results at only a marginal improvement in the statistic ($\Delta \chi^2=6.7$ for two additional dofs) 
in $E_\mathrm{line}=14.65_{-0.69}^{+0.86}$~MeV and the total photon flux in the line of $5.9_{-2.3}^{+2.4}$~ph~cm$^{-2}$~s$^{-1}$, 
which implies the line isotropic luminosity  $L_\mathrm{line}\sim 9.2 \pm 3.8 \times10^{51}$~erg~s$^{-1}$.
We estimate the improvement significance by applying the Akaike Information Criterion (see, e.g., \citealt{Burnham_2004}), 
the method employed by R23, and find that the addition of the Gaussian line to the model results in only a small decrease in the value of the criterion $\Delta \rm{AIC}\approx2.7$ that corresponds to $<1\ \sigma$ significance of the improvement. 

Nevertheless, we note that the estimated line central energy fits well the decaying trend of $E_\mathrm{line}$ reported in R23, 
and the implied ratio $L_\mathrm{line,51}/(E_\mathrm{line}/\rm{MeV})\sim0.63$ is in the range, predicted at times close to the emission peak 
by one of the emission scenarios explored by the authors, which involves high-latitude emission (HLE) from the shell that produced the most luminous pulse in the GRB light curve.

\begin{acknowledgments}
This work is supported by RSF grant 21-12-00250. 
S.M. and A.A.L. acknowledge the support by the RFBR grant 19-29-11029 in the part of the ART-XC data analysis.
This work made use of data supplied by the UK Swift Science Data Centre at the University of Leicester.

\end{acknowledgments}

\vspace{5mm}
\facilities{\textit{Wind} (Konus), \textit{SRG} (ART-XC)}

\appendix

\section{\KW\ Data Corrections} \label{app:corr}

\subsection{Light Curves}  \label{app:corrLC}
A standard Konus-Wind dead-time (DT) correction procedure for light curves 
is a simple non-paralyzable DT correction in each of the measurement bands, 
with a dead time $\tau$ of $\sim 4\mu$s, taking into account a softer gate blocking by harder ones. 
This method, based on a relation $1/n+1/N=\tau$ between the total photon 
flux $N$ incident on the detector (assuming 100\% detection efficiency) and the recorded count rate $n$, provides a robust flux estimate for $N\lesssim 1/\tau$ ($\sim 2.5\times10^{5}$~cts~s$^{-1}$). At $N\gg 1/\tau$ $n$ ceases to depend on $N$ (saturates) and the standard DT correction becomes ineffective.

At very high incident fluxes, pulse pileups in detector electronics lead to multiple analog and digital distortion effects that require special efforts to correctly reconstruct the time history of the event and the energy spectra. The instrument response to fluxes up to $10^{6}-10^{7}$~cts~s$^{-1}$ of various incident photon spectra was studied in laboratory experiments with strong radioactive and X-ray sources, as well as in Monte Carlo simulations of KW analog and digital electronics behavior \citep{Mazets_1999AstL...25..635M}.
It was found that the pattern of the pileup-distorted $n(N)$ relations for each of the three energy bands G1, G2, and G3 is strongly sensitive to the shape of the incident photon spectrum. 
Hence, by comparing the behavior of observed rates $g_{1,2,3}$ (in G1, G2, and G3, respectively ) with the $n_{1,2,3}(N)$ dependencies obtained from simulations for different energy spectra, one can reliably reconstruct not only the incident emission intensity but also its spectral shape. Based on this approach, deconvolution procedures were developed that allowed, e.g., to successfully recover time histories and energy spectra of extremely bright magnetar flares \citep{Mazets_1999ApJ...519L.151M,Mazets_1999AstL...25..635M}.   

To reconstruct \GRB\ light curves in the triggered detector S2 at the peak of the emission, 
we, following the approach of \citep{Mazets_1999AstL...25..635M}, performed Monte Carlo simulations 
for various incident count fluxes $N$ and Band-shaped photon spectra, forward-folded with the detector response matrix. 
From the simulations, we obtained a database of $\sim 10,000$ $n_i(N,\alpha,\beta,E_p)$ dependencies 
for $N$ up to $\sim 4\times10^{7}$~s$^{-1}$, $\alpha$ in range $(-1.8, +1.0$), $\beta$ in range $(-3.0,-2.0)$, and $E_p$ in the range from 500~keV to 4.5~MeV.

Then, for each time bin in the interval from 216~s to 250~s, 
we searched the database for the best match of a simulated triplet $n_{1,2,3}(N,\alpha,\beta,E_p)$ and the measured rates $g_{1,2,3}(t)$ by minimizing the sum in quadrature of normalized differences $(n_i(N,\alpha,\beta,E_p)-g_{i}(t))/g_{i}(t)$.
In order to reduce the number of free spectral parameters in the search, $\beta(t)$ was fixed 
to that obtained from a preliminary fit to the corresponding multichannel spectrum in the PHA2 band (0.4--16.5~MeV); 
this approach is justified by the fact that the hard end of the KW instrumental spectrum (at energies above $\sim 2-4$~MeV) 
remains nearly undistorted by pileups, and hence, the high-energy spectral index can be estimated independently. 
In a case of ambiguous identification, we manually selected the best-solution parameters ($N$, $\alpha$, $E_p$) using the following criteria. 
First, the variation of the emission intensity ($N$) over time had to follow its general course in the second KW detector S1, 
for which, for this GRB, saturation and pileup effects are negligible due to the emission absorption in the \textit{Wind} body and the rear structure of the detector. Second, we aimed to achieve smooth variations in the spectral parameters $\alpha$ and $E_p$ over time. 
Finally, using the best-solution parameters ($\alpha(t),\beta(t),E_p(t)$), we calculated a Band-shaped spectrum, normalized on the incident count flux $N(t)$ in the whole instrument energy range, and reconstructed count rates Gi$(t)$ were calculated, from this spectrum, as count fluxes in the corresponding energy band Gi. 

As a result, we obtained the reconstructed incident count rates as well as time-resolved estimates 
of the spectral parameters (illustrated in Figure~\ref{FigP2P3}). To estimate uncertainties in these values, 
we performed simulations by varying (assuming Poisson-distributed counts) the measured rates for several time bins. 
The resulting variations in the best-solution parameters do not exceed $\sim 0.15$ for $\alpha$, $\sim 10$\% for $E_p$, and about 18\% for the flux $N$.  
 
\subsection{Multichannel Spectra}  \label{app:corrPHA}

A standard Konus-Wind dead-time correction procedure for multichannel spectra is 
similar to that for the light curves, but with about 10 times longer $\tau\approx42\mu$s.
Accordingly, count rate saturations in PHA1 and PHA2 are not negligible 
at $N\gtrsim 2.4\times10^{4}$~s$^{-1}$, and pileup corrections become necessary 
at $N\gtrsim 5\times10^{4}$~s$^{-1}$ (in the corresponding spectral band).
The influence of the pileup effect on KW spectra was examined in studies 
of powerful solar flares \citep{Lysenko_2019ApJ...877..145L,Lysenko_2022ApJS..262...32L}.
An iterative correction method for pileup-distorted spectra was developed that allowed, e.g., 
to recover, at incident count rates up to $\sim 2\times10^{5}$~cts~s$^{-1}$,
steep, broken power-law spectral shapes to the accuracy 
of $\sim 0.1$ in the spectral indices and $\sim 10$~keV 
in the break energy. In these studies, spectral shape corrections were applied to PHA1 (20--1200~keV), 
whereas the flux was corrected using joint spectral fits with nearly undistorted and unsaturated 
spectra in the PHA2 range (0.4--16.5~MeV).   

For \GRB\,, both saturation and pileup corrections are necessary 
to three 8.192-s long spectra measured from 216.832~s to 241.408~s after the trigger. 
In this time interval, a huge incident flux in the PHA1 band (up to millions counts/s) 
makes corrections with a method similar to that of \cite{Lysenko_2019ApJ...877..145L} very difficult, if not impossible.
However, such a procedure is still applicable to spectra in the PHA2 band, where incident rates do not exceed $\sim 3\times10^{5}$~cts~s$^{-1}$,
but an ``external'' normalization is still required to correct the deeply saturated measured flux. The flux corrections were performed by simultaneous fits of the shape-corrected
PHA2 spectra with three spectral points in the 20--1200~keV band, 
constructed from the light-curve data, which were corrected independently (see Appendix~\ref{app:corrLC}).

For the spectrum at the peak of the emission (225.024--233.216~s after the trigger), 
an additional spectral point is available from the unsaturated Z-channel data (16.5--22~MeV),
thus providing an independent reference at the higher energies. 
For this spectrum, joint fits were made to three data sets: 
THA+PHA2; PHA2+Z; and THA+PHA2+Z. The fits result in very similar 
spectral parameters and fluxes, which confirms the correctness 
of our approach to recover both the pileup-distorted shape of the spectrum 
and the saturated incident flux.

\bibliography{grb221009a}{}
\bibliographystyle{aasjournal}

\end{document}